\documentstyle{article}

\begin{document}
\baselineskip=18pt

\title{\hspace{11cm} {\small \bf IMPNW-970203 }\\
\vspace{2cm}Bosonization of vertex operators 
for Zn symmetric Belavin model and its correlation functions}

\author{
0\bigskip\\
$^{a}$ CCAST (World Laboratory), P.O.Box 8730, Beijing 100080, China\\
$^{b}$ Institute of Modern Physics, Northwest 
University, Xian 710069, China
\thanks{Mailing address}}
\date{February 3,1997}
\maketitle

\begin{abstract}
Based on the bosonization of vertex operators for $A_{n-1}^{(1)}$ face model 
by Asai,Jimbo, Miwa and Pugai,using vertex-face correspondence we obtain 
vertex operators for Zn symmetric Belavin model,which are constructed 
by deformed boson oscilllators. The correlation functions are also obtained.

\end{abstract}

\section{Introduction}
The Bosonization of vertex operators for solvable models is very powerful 
for studying their correlation functions$^{[1-3]}$.These vertex operators 
realize the Zamolodichikov-Fadeev algebras$^{[4-7]}$ with the R-matrices of 
the models.From the bosonized vertex operators,one can obtain  
multi-point correlation functions as explicit integrals.

Recently,the studies of q-deformed Virasoro algebra$^{[8,21,22 ]}$ 
and its vertex 
operators make it clear to understand  the correspondence between the 
unitary minimal conformal models$^{[26]}$ and the ABF models$^{11]}$,
this thing has 
been considered as a mysterious one for a long time.The bosonization 
for q-deformed 
Virasoro algebra and its vertex operators also make  it possible 
to calculate the 
correlation functions for ABF models.The bosonization for q-deformed 
W algebra$^{[21,22]}$ and its vertex operations$^{[10,21]}$ 
would promote  to 
investigate the $A^{(1)}_{n}$ RSOS models$^{[15]}$.Naively,the q-defomed W 
algebra 
(including q-deformed Virasoro algebra) would play an important role in 
the elliptic face models.How about the elliptic vertex model---eight vertex 
model and Zn Belavin model ? Jimbo et al obtained the difference equations 
for 
eight-vertex operators using the method of  the corner transfer matrix (CTM) 
and  ``physical picture", morever, give the spontaneous polarization of 
eight-vertex 
model$^{[20]}$.Through the same method, Quano obtained the 
difference equations for 
Zn Belavin model and the corresponding spontaneous polarization$^{[19]}$. 
However ,to calculate the  
general correlation functions for the elliptic vertex model practically is 
very complicated and still is an open problem$^{[24,27]}$.

After the works for trigonomatric models$^{[1-3]}$ several important works 
for elliptic models has been done$^{[8-10]}$.Lukyanov et al  give the 
bosonization
of vertex operators for ABF model.Miwa give the corresponding bosonized 
boundary 
operators $^{[9]}$ .Recently,Asai,Jimbo  et al obtain the bosonized vertex 
operators $^{10]}$ for $A^{(1)}_{n}$ face model$^{[12]}$.These work will 
greatly promote the study of solvable models of elliptic type. Based on 
their work, using vertex-face correspondence $^{[13-15]}$,we obtain 
bosonized vertex operators 
for Zn symmetric Belavin model$^{[16,17]}$.

In the following ,we first find an intertwiner in section 2,3, which 
intertwines the Belavin R-matrix and the Boltzmann weight of 
$A^{(1)}_{n}$ face model in 
Ref.[10]. We then review the vertex operators in Ref.[8-10] and their 
exchange relations in section 4.Combining their vertex operators and the 
intertwiners ,  we finally obtain the bosonization for vertex operators 
 and give the correlation functions of Zn symmetric 
Belavin model in section 5.

\section{Vertex-face correspondence}
Given a integer n ($2\leq n$),$\tau$ ,w$\in C$ and Im$\tau >0$ ,we can 
construct Zn symmetric Belavin R-matrix$^{[16,17]}$. Define $n\times n$ 
 matrices $g,h ,I_{\alpha}$ where
\begin{eqnarray*}
& &g_{jk}=\omega^{j}\delta_{jk}\ \ ,\ \ h_{jk}=\delta_{j+1,k},\ \ 
\omega =exp(\frac{2i\pi}{n})\\
& &I_{\alpha}=I_{(\alpha_{1},\alpha_{2})}=g^{\alpha_{2}}h^{\alpha_{1}}\ \ ,
\ \ (\alpha_{1},\alpha_{2})\in Z_{n}^{2}
\end{eqnarray*}
\noindent Define $I_{\alpha}^{(j)}=I\otimes I\otimes ....\otimes I_{\alpha}\otimes I
\otimes....\otimes I$ ,where $I_{\alpha}$ is at the $j^{th}$ site, I is the 
$n\times n $ unit matrix,and  
\begin{eqnarray*}
& &W_{\alpha}(z,\tau)=\frac{1}{n}
\theta\left[\begin{array}{l}\frac{1}{2}+
\frac{\alpha_{2}}{n}\\ \frac{1}{2}+\frac{\alpha_{1}}{n}\end{array}\right]
(z+\frac{w}{n},\tau)/\theta\left[\begin{array}{l}\frac{1}{2}+
\frac{\alpha_{2}}{n}\\ \frac{1}{2}+\frac{\alpha_{1}}{n}\end{array}\right]
(\frac{w}{n},\tau)\\
& & \theta\left[\begin{array}{l}a\\b\end{array}\right]
(z,\tau)=\sum_{m\in Z }exp\{i\pi(m+a)[(m+a)\tau +2(z+b)]\}\\
& &\sigma_{0}(z,\tau)=\theta\left[\begin{array}{l}\frac{1}{2}\\
\frac{1}{2}\end{array}\right](z,\tau)
\end{eqnarray*}
\noindent Zn symmetric Belavin R-matrix is 
\begin{eqnarray}
R_{jk}(z,\tau)=\frac{\sigma_{0}(w,\tau)}{\sigma_{0}(z+w,\tau)}
\sum_{\alpha\in Z_{n}^{2}}W_{\alpha}(z,\tau)I_{\alpha}^{(j)}
(I_{\alpha}^{-1})^{(k)}
\end{eqnarray}
\noindent which satisfies Yang-Baxter equation (YBE)
\begin{eqnarray}
& &R_{12}(z_{1}-z_{2},\tau) R_{13}(z_{1}-z_{3},\tau) R_{23}(z_{2}-z_{3},\tau)
=R_{23}(z_{2}-z_{3},\tau)R_{13}(z_{1}-z_{3},\tau)R_{12}(z_{1}-z_{2},\tau)
\end{eqnarray}

Given a n-vector $a\in Z^{n}$ , we define Boltzmann weight of $A_{n-1}^{(1)}$ 
face model$^{[12]}$,which can be written in the vertex form $W(a|z,\tau)
^{\mu'\nu'}_{\mu\nu}$, where the non-zero elements 
are 
\begin{eqnarray}
& &W(a|z,\tau)^{\mu\mu}_{\mu\mu}=1\\
& &W(a|z,\tau)^{\mu\nu}_{\nu\mu}=
\frac{\sigma_{0}(z+w,\tau)}{\sigma_{0}(w,\tau)}
\frac{\sigma_{0}(z+a_{\mu\nu}w,\tau)}
{\sigma_{0}(a_{\mu\nu}w,\tau)}\ \ ,\ \ \mu\neq\nu\\
& &W(a|z,\tau)^{\mu\nu}_{\mu\nu}=\frac{\sigma_{0}(z,\tau)
\sigma_{0}(a_{\mu\nu}w-w,\tau)}{\sigma_{0}(z+w,\tau)
\sigma_{0}(a_{\mu\nu}w,\tau)}\ \  \ , \mu\neq\nu
\end{eqnarray}
\noindent where $a_{\mu\nu}$ is defined by $a=(a_{1},....,a_{n})$ by
\begin{eqnarray}
& &\overline{a}_{\mu}=a_{\mu}-\frac{1}{n}\sum_{l=1}^{n}a_{l}+w_{\mu}\\
& &\overline{a}_{\mu\nu}=\overline{a}_{\mu}-\overline{a}_{\nu}=
a_{\mu}-a_{\nu}+w_{\mu}-w_{\nu}
\end{eqnarray}
\noindent $\{w_{j}\}$ is a set of generic complex numbers specified by the 
face model under investigation.

The intertwiners of vertex-face correspondence are $^{[14,15]}$ n-column 
vectors $\varphi_{\mu ,a}(z,\tau)$ whose k-th component is 
\begin{eqnarray*}
& &\varphi_{\mu ,a}^{(k)}(z,\tau)=\theta^{(j)}(z+
nw(\overline{a}_{\mu}+1-\frac{1}{n}),\tau)\\
& &\theta^{(j)}(z,\tau)=\theta\left[\begin{array}{l}\frac{1}{2}-\frac{j}{n}\\
 \frac{1}{2}\end{array}\right](z,n\tau)
 \end{eqnarray*}
\noindent The vertex-face correspondence is 
\begin{eqnarray}
R(z_{1}-z_{2},\tau)\varphi_{\mu ,a+e_{\nu}}(z_{1},\tau)\otimes 
\varphi_{\nu ,a}(z_{2},\tau)
=\sum_{\mu '\nu '}W(a|z_{1}-z_{2},\tau)^{\mu\nu}_{\mu '\nu '}
\varphi_{\mu ' ,a}(z_{1},\tau)\otimes \varphi_{\nu ',a+e_{\mu '}}
(z_{2},\tau)
\end{eqnarray}
\noindent where $e_{\mu}=(0,0,....,1,0,...,0)$ and $`` 1"$ is at the 
$\mu^{th}$ site. We can introduce n-row vectors $\stackrel{\sim}{\varphi}
_{\mu ,a}(z,\tau)$ such that 
\begin{eqnarray}
\sum_{k}\stackrel{\sim}{\varphi}^{(k)}_{\mu ,a}(z,\tau)
\varphi^{(k)}_{\nu ,a}(z,\tau)
=\delta_{\mu\nu}
\end{eqnarray}
\noindent Thus we have 
\begin{eqnarray}
\sum_{\mu}\varphi_{\mu ,a}(z,\tau)\stackrel{\sim}{\varphi}_{\mu ,a}
(z,\tau)=I
\end{eqnarray}
\noindent Notice $\stackrel{\sim}{\varphi}^{(k)}_{\mu ,a}(z,\tau)$ 
is a function 
of $a,\mu ,k,z,n,\tau ,w,\{w_{j}\}$. One can show the vertex-face 
correspondence by $\stackrel{\sim}{\varphi}$  
\begin{eqnarray}
\stackrel{\sim}{\varphi}_{\mu ,a}(z_{1},\tau)\otimes 
\stackrel{\sim}{\varphi}
_{\nu ,a+e_{\mu}}(z_{2},\tau)R(z_{1}-z_{2},\tau)=
\sum_{\mu '\nu '}W(a|z_{1}-z_{2},\tau)^{\mu '\nu '}_{\mu\nu}
\stackrel{\sim}{\varphi}_{\mu ',a+e_{\nu '}}(z_{1},\tau)\otimes 
\stackrel{\sim}{\varphi}_{\nu ',a}(z_{2},\tau)
\end{eqnarray}
\section{Modular transformation of Boltzmann weight}

Since the Boltzmann weight in Eq.(3)-Eq.(5) is not the same as that 
in Ref.[10]. We
need to rescale the Zn Belavin R-matrix and sepcify the parameters $z$ 
$\tau$ , so that we could directly use the beautiful results in Ref.[10].
Let us restrict the parameter $w$ : Im$w>0$ and set
\begin{eqnarray*}
& &x=e^{i\pi w}\ \ \ ,\ \ |x|<1\ \ \ \ ({\rm Im}w>0)\ \ ,\ \ n+2\leq r\\
& &\left[v\right]=e^{i\pi\frac{wv^{2}}{r}}\sigma_{0}(wv,rw)=const\times 
x^{\frac{v^{2}}{r}-v}\Theta_{x^{2r}}(x^{2v})\\
& &\Theta_{q}(z)=(z,q)(qz^{-1},q)(q,q)\ \ \ \ ,\ \ (z,q)=\prod_{n=0}^{\infty}
(1-zq^{n})
\end{eqnarray*}
\noindent Notice the modular transformation for theta function 
$\sigma_{0}(z,\tau)$
\begin{eqnarray}
\sigma_{0}(\frac{z}{\tau},-\frac{1}{\tau})=const\times 
e^{\frac{i\pi z^{2}}{\tau}}\sigma_{0}(z,\tau)
\end{eqnarray}
\noindent we have
\begin{eqnarray}
\left[v\right]=const \sigma_{0}(\frac{vw}{rw},-\frac{1}{rw})
\end{eqnarray}
Let Zn symmetric Belavin R-matrix be rescaled, and the parameter $z$  
$\tau$ be specified as follows
\begin{eqnarray}
& &R(v,-\frac{1}{rw})=r_{1}(-v)\frac{\sigma_{0}(\frac{1}{r},-\frac{1}{rw})}
{\sigma_{0}(\frac{v+1}{r},-\frac{1}{rw})}\sum_{\alpha}W_{\alpha}
(v,-\frac{1}{rw})I_{\alpha}\otimes I_{\alpha}^{-1}\\
& &W_{\alpha}(v,-\frac{1}{rw})=\frac{
\sigma_{\alpha}(\frac{v}{r}+\frac{1}{nr},-\frac{1}{rw})}
{n\sigma_{\alpha}(\frac{1}{nr},-\frac{1}{rw})} \nonumber\\
& &r_{1}(v)=x^{2v\frac{(r-1)(n-1)}{nr}}\frac{g_{1}(-v)}{g_{1}(v)}\ \ ,\ \ 
g(v)=\frac{\{x^{2+2v}\}\{x^{2r+2n-2+2v}\}}
{\{x^{2r+2v}\}\{x^{2n+2v}\}}\nonumber\\
& &\{z\}=(z;x^{2r},x^{2n})\ \ \ ,\ \ \ 
(z;q_{1},q_{2},...,q_{m})=\prod^{\infty}_{\{n_{j}\}=0}(1-zq_{1}^{n_{1}}
q_{2}^{n_{2}}...q_{m}^{n_{m}})\nonumber
\end{eqnarray}
\noindent we also specify the intertwiners $\varphi$ and 
$\stackrel{\sim}{\varphi}$ as follows
\begin{eqnarray}
& &\varphi^{(k)}_{\mu ,a}(v,-\frac{1}{rw})=\theta^{(k)}(\frac{v+
n(\overline{a}_{\mu}+1-\frac{1}{n})}{r},-\frac{1}{rw})\nonumber\\
&&\sum_{k}\stackrel{\sim}{\varphi}^{(k)}_{\mu ,a}(v,-\frac{1}{rw})
\varphi^{(k)}_{\nu ,a}(v,-\frac{1}{rw})
=\delta_{\mu\nu}
\end{eqnarray}
\noindent The vertex-face correspondence becomes 
\begin{eqnarray}
& &\stackrel{\sim}{\varphi}_{\mu ,a}(v_{1},-\frac{1}{rw})\otimes 
\stackrel{\sim}{\varphi}
_{\nu ,a+e_{\mu}}(v_{2},-\frac{1}{rw})R(v_{1}-v_{2},-\frac{1}{rw})\nonumber\\
& &\ \ \ =\sum_{\mu '\nu '}W'(a|v_{1}-v_{2},
-\frac{1}{rw})
^{\mu '\nu '}_{\mu\nu}
\stackrel{\sim}{\varphi}_{\mu ',a+e_{\nu '}}(v_{1},-\frac{1}{rw})\otimes 
\stackrel{\sim}{\varphi}_{\nu ',a}(v_{2},-\frac{1}{rw})
\end{eqnarray}
\noindent using the Eq.(11), the non-zero element of 
$W'(a|v,-\frac{1}{rw})^{\mu '\nu '}_{\mu\nu}$ can be written  
\begin{eqnarray}
& &W'(a|v,-\frac{1}{rw})^{\mu\mu}_{\mu\mu}=r_{1}(-v)\\
& &W'(a|v,-\frac{1}{rw})^{\mu\nu}_{\mu\nu}=r_{1}(-v)\frac
{\left[v\right]\left[a_{\mu\nu}-1\right]}
{\left[v+1\right]\left[a_{\mu\nu}\right]}\\
& &W'(a|v,-\frac{1}{rw})^{\mu\nu}_{\nu\mu}=r_{1}(-v)\frac
{\left[v+a_{\mu\nu}\right]\left[1\right]}
{\left[v+1\right]\left[a_{\mu\nu}\right]}
\end{eqnarray}
It can be found that our Boltzmann weight 
$W'(a|-v,-\frac{1}{rw})^{\mu '\nu '}_{\mu\nu}$ is the same as the Boltzmann 
weight $W\left(\begin{array}{ll}a+\overline{\epsilon}_{\mu}
+\overline{\epsilon}_{\nu}&
a+\overline{\epsilon}_{\mu}\\a+\overline{\epsilon}_{\nu}&a\end{array}|v
\right)$ in  the Ref.[10]
\begin{eqnarray}
& &W'(a|-v,-\frac{1}{rw})^{\mu\mu}_{\mu\mu}
=r_{1}(v)=W\left(\begin{array}{ll}a+2\overline{\epsilon}_{\mu}&
a+\overline{\epsilon}_{\mu}\\a+\overline{\epsilon}_{\mu}&a\end{array}|v
\right)
\\
& &W'(a|-v,-\frac{1}{rw})^{\mu\nu}_{\mu\nu}
=r_{1}(v)\frac
{\left[v\right]\left[a_{\mu\nu}-1\right]}
{\left[v-1\right]\left[a_{\mu\nu}\right]}\nonumber\\
& &\ \ =W\left(\begin{array}{ll}a+\overline{\epsilon}_{\mu}
+\overline{\epsilon}_{\nu}&
a+\overline{\epsilon}_{\mu}\\a+\overline{\epsilon}_{\nu}&a\end{array}|v
\right)
\\
& &W'(a|-v,-\frac{1}{rw})^{\mu\nu}_{\nu\mu}=r_{1}(v)\frac
{\left[v-a_{\mu\nu}\right]\left[1\right]}
{\left[v-1\right]\left[a_{\mu\nu}\right]}\nonumber\\
& &\ \ =W\left(\begin{array}{ll}a+\overline{\epsilon}_{\mu}
+\overline{\epsilon}_{\nu}&
a+\overline{\epsilon}_{\nu}\\a+\overline{\epsilon}_{\nu}&a\end{array}|v
\right)
\end{eqnarray}
\section{Vertex operators in $A^{(1)}_{n-1}$ face model}
We review Asai,Jimbo,Miwa and Pugai's bosonization of vertex operator  
 for $A^{(1)}_{n-1}$ face model$^{[10]}$.

According to Ref.[8-10],introduce bosonic oscillators $\beta^{j}_{m}$ (
$1\leq j\leq n-1 ,m\in Z\backslash \{0\}$) which satisfy
\begin{eqnarray}
& &[\beta^{j}_{m},\beta^{j}_{m'}]=m\frac{[(n-1)m]_{x}[(r-1)m]_{x}}
{[nm]_{x}[rm]_{x}}\delta_{m+m',0}\\
& &[\beta^{j}_{m},\beta^{k}_{m'}=-mx^{{\rm sgn}(j-k)nm}\frac{[m]_{x}
[(r-1)m]_{x}}{[nm]_{x}[rm]_{x}}\delta_{m+m',0}\ \ ,\ \ j\neq k\\
& &[a]_{x}=\frac{x^{a}-x^{-a}}{x-x^{-1}}\ \ \ ,\ \ x=e^{i\pi w}\nonumber
\end{eqnarray}
\noindent Define $\beta^{n}_{m}=-x^{2mn}\sum^{n-1}_{j=1}x^{-2jm}
\beta^{j}_{m}$ . 

Introduce zero modes $p_{\mu}$, $q_{\mu} \ \ (\mu =1,...,n)$ , such that 
$[ip_{\mu},q_{\nu}]=\delta_{\mu ,\nu}$ .Consider orthnormal bases 
$\{e_{\mu}\} \ \ ,\mu =1,....,n$ $<e_{\mu},e_{\nu}>=\delta_{\mu\nu}$, and 
\begin{eqnarray*}
\overline{e}_{\mu}=e_{\mu}-\frac{1}{n}\sum_{k}e_{k}
\end{eqnarray*}
\noindent Define 
\begin{eqnarray*}
Q_{\overline{e}_{\mu}}=q_{\mu}\ \ \ ,\ \ P_{\overline{e}_{\mu}}=
p_{\mu}-\frac{1}{n}\sum_{k}p_{k}
\end{eqnarray*}
\noindent One have 
\begin{eqnarray*}
[iP_{\overline{e}_{\mu}},Q_{\overline{e}_{\nu}}]=\delta_{\mu\nu}-\frac{1}{n}
=<\overline{e}_{\mu},\overline{e}_{\nu}>
\end{eqnarray*}
\noindent Let the vacuum $|0>$ be such that
\begin{eqnarray*}
\beta^{j}_{m}|0>=p_{\mu}|0>=0\ \ ,\ \ {\rm for} \ \ m>0
\end{eqnarray*}
\noindent and that
\begin{eqnarray*}
|l,k>=e^{i\sqrt{\frac{r}{r-1}}Q_{l}-i\sqrt{\frac{r-1}{r}}Q_{k}}|0>
\end{eqnarray*}
\noindent where $l=\sum_{j=1}l_{j}e_{j}$  $k=\sum_{j=1}k_{j}e_{j}$  
$Q_{k}=\sum_{j}k_{j}q_{j}$ $ Q_{l}=\sum_{j}l_{j}q_{j}$ ,and let
\begin{eqnarray*}
\gamma=\gamma_{j}e_{j}\ \ ,\ \  \beta=\sum_{j}\beta_{j}e_{j}\ \ ,\ \ 
P_{k}=\sum_{j}k_{j}P_{\overline{e}_{j}}
\end{eqnarray*}
\noindent we have 
\begin{eqnarray*}
[iP_{\gamma},Q_{\beta}]=<\overline{\gamma},\beta>=<\gamma,\overline{\beta}>
\end{eqnarray*}
\noindent where $\overline{\gamma}=\sum_{j}\gamma_{j}\overline{e}_{j}$ 
$\overline{\beta}=\sum_{j}\beta_{j}\overline{e}_{j}$. The Fock space 
$F_{l,k}=C[\{\beta^{j}_{-1},\beta^{j}_{-2},...\}_{1\leq j\leq n}]|l,k>$ with 
\begin{eqnarray}
& &\beta^{j}_{m}|l,k>=0\ \ (m>0)\nonumber\\
& &P_{\gamma}|l,k>=<\overline{\gamma},\sqrt{\frac{r}{r-1}}l-
\sqrt{\frac{r-1}{r}}k>
\end{eqnarray}
Define for j=1,...,n-1
\begin{eqnarray*}
& &{\rm the\ \ simple\ \ root\ \ } \alpha_{j}=e_{j}-e_{j+1}\ \ \ ,\ \ \ 
{\rm the \ \ basic\ \  weight\ \ } \omega_{j}=
\sum_{k=1}^{j}\overline{e}_{k}\\
& &\xi_{j}(v)=
e^{i\sqrt{\frac{r-1}{r}}(Q_{\alpha_{j}}-iP_{\alpha_{j}}2vlnx)}
:e^{\sum_{m\neq 0}\frac{1}{m}(\beta^{j}_{m}-\beta^{j+1}_{m})x^{-(j+2v)m}}:\\
& &\eta_{j}(v)=
e^{-i\sqrt{\frac{r-1}{r}}(Q_{\omega_{j}}-iP_{\omega_{j}}2vlnx)}
:e^{-\sum_{m\neq 0}\frac{1}{m}\sum^{j}_{k=1}\beta^{k}_{m}x^{(j-2k+1-2v)m}}:
\end{eqnarray*}
Introduce vertex operators
\begin{eqnarray}
& &\phi_{\mu}(v)=\oint\prod^{\mu-1}_{j=1}\frac{d(x^{2v_{j}})}{2i\pi 
x^{2v_{j}}}
\eta_{1}(v)\xi_{1}(v_{1})...\xi_{\mu-1}(v_{\mu-1})\prod^{\mu-1}_{j=1}
f(v_{j}-v_{j-1},\hat{\pi}_{j\mu})\\
& &\hat{\pi}_{\mu}=\sqrt{r(r-1)}P_{\overline{e}_{\mu}}+w_{\mu}\ \ 
f(v,y)=\frac{[v+\frac{1}{2}-y]}{[v-\frac{1}{2}]}\nonumber
\end{eqnarray}
\noindent where $\{w_{j}\}$ is a set complex number defined in Eq.(6) 
and  $\hat{\pi}_{\mu}$ is a set of operators.Here we set $v_{0}=v$ and 
take the integration contours to be simple closed curves around the 
origin satisfying
\begin{eqnarray*}
|xx^{2v_{j-1}}|<|x^{2v_{j}}|<|x^{-1}x^{2v_{j-1}}|\ \ \ (j=1,....,\mu -1)
\end{eqnarray*}
Following Ref.[10],one can verify
\begin{eqnarray*}
\phi_{\mu}(v_{1})\phi_{\nu}(v_{2})=\sum_{\mu '\nu '}\phi_{\nu '}(v_{2})
\phi_{\mu '}(v_{1})W'(\hat{\pi}|v_{2}-v_{1},-\frac{1}{rw})
^{\mu\nu}_{\mu '\nu '}
\end{eqnarray*}
\noindent Namely,
\begin{eqnarray}
\phi_{\mu}(-v_{1})\phi_{\nu}(-v_{2})=\sum_{\mu '\nu '}\phi_{\nu '}(v_{2})
\phi_{\mu '}(v_{1})
W'(\hat{\pi}|v_{1}-v_{2},-\frac{1}{rw})
^{\mu\nu}_{\mu '\nu '}
\end{eqnarray}
\noindent Now the Boltzmann weight $W'(\hat{\pi}|v,-\frac{1}{rw})
^{\mu\nu}_{\mu '\nu '}$ be some functions like Eq.(17)-Eq.(19) with 
$a_{\mu\nu}$ replaced by operator $\hat{\pi}_{\mu\nu}$.Thus it does not  
commutate with vertex operator $\phi_{\mu}(v)$ and the exchange relations 
 Eq.(27) should be written in that order.

\section{vertex operators for Zn symmetric Belavin model and its correlation 
functions}
In ``physical picture" of lattice models$^{[1,3,15,20]}$ ,the vertex 
operators for 
Zn symmetric Belavin model can be realized by a half column transfer matrix, 
and these vertex operators realize the Zamolodchikov-Fadeev algebra with 
the Zn symmetric R-matrix as its construction coefficent$^{[19,20,29]}$.
\begin{eqnarray}
Z^{j}(z_{1})Z^{k}(z_{2})=\sum_{j'k'}Z^{k'}(z_{2})Z^{j'}(z_{1})R^{jk}_{j'k'}
(z_{1}-z_{2},\tau)
\end{eqnarray}
\noindent where $Z^{j}(z)$ are some operators acting on the eigenvaluevectors 
spaces ${\bf H}^{i}$ of the Hamiltonian $D^{(i)}$ of CTM ,where 
\begin{eqnarray}
e^{i\pi wnD}Z^{j}(z)e^{-i\pi wnD}=Z^{j}(nw+z)
\end{eqnarray}
\noindent The correlation 
 functions are expressed in terms of the trace of vertex operators in the 
spaces ${\bf H}^{i}$. For the detail, we refer readers to the Ref.[3].

Our main idea is to realize these vertex operators and the $D^{(i)}$ in a 
direct sum of Fock 
space $L_{i}=\oplus_{\{m_{i}\}\in Z}F_{l,\overline{\Lambda}_{i}+
\sum_{j=1}^{n-1}m_{j}\alpha_{j}}$ ,$\overline{\Lambda}_{i}$ ($\alpha_{i}$)  
is the basic weight of Lie algebra $A_{n-1}$ (resp. simple root of $A_{n-1}$) 
This representation is expect to be reducible.In fact ,the $H^{i}$ can be  
consider as the same as the Fock space $L_{i}$ from the chacter $^{[18]}$ 
(see Eq.(39)) 
\begin{eqnarray}
tr_{H^{i}}(x^{nD^{(i)}})=tr_{L_{i}}(x^{nD^{(i)}})
\end{eqnarray}
\noindent A simlar bosonization produce was applied for caculations of 
conformal blocks in the conformal field theory (CTF)$^{[28]}$ , lattice 
correlation functions for XXZ model$^{[3]}$ and ABF model$^{[8,9]}$.

Define $\hat{a}_{\mu}=-\sqrt{\frac{r}{r-1}}p_{\mu}+w_{\mu}
+\frac{r}{r-1}<e_{\mu},l>\ \ \hat{a}_{\mu\nu}=\hat{a}_{\mu}-\hat{a}_{\nu}$ 
,we have 
\begin{eqnarray}
& &\hat{a}_{\mu}e^{-i\sqrt\frac{r-1}{r}Q_{\nu}}=e^{-i\sqrt\frac{r-1}{r}
Q_{\nu}}(
\hat{a}_{\mu}+\delta_{\mu\nu})\nonumber\\
& &\hat{\pi}_{\mu\nu}F_{l,k}=(r<e_{\mu}-e_{\nu},l-k>+<e_{\mu}-e_{\nu},k>+
w_{\mu}-w_{\nu})F_{l,k}\nonumber\\
& &\hat{a}_{\mu\nu}F_{l,k}=(<e_{\mu}-e_{\nu},k>+
w_{\mu}-w_{\nu})F_{l,k}
\end{eqnarray}
\noindent From above equation and $[v+r]=-[v]$, we can derive 
\begin{eqnarray}
W'(\hat{\pi}|v,-\frac{1}{rw})
^{\mu\nu}_{\mu '\nu '}|_{F_{l,k}}=
W'(\hat{a}|v,-\frac{1}{rw})
^{\mu\nu}_{\mu '\nu '}|_{F_{l,k}}
\end{eqnarray}

From Eq.(11) and Eq.(27), we can construct the bosonization for 
the vertex operators of Zn symmetric Belavin model which satisfy 
relation Eq(28) with sepcifying the parameters $z$ , $\tau$.  
Define
\begin{eqnarray}
\Phi^{(j)} (v)=\sum_{\mu}\phi_{\mu}(-v)\stackrel{\sim}
{\varphi}_{\mu ,\hat{a}}^{(j)}(v+\delta,-\frac{1}{rw})
\end{eqnarray}
\noindent where $\delta$ is a generic parameter. 
Notice Eq.(27) , Eq.(33) and the vertex-face correspondence Eq(11), 
we have 
\begin{eqnarray*}
& &\Phi^{(i)}(v_{1})\Phi^{(j)}(v_{2})|_{F_{l,k}}=
\sum_{\mu\nu}\phi_{\mu}(-v_{1})\stackrel{\sim}{\varphi}^{(i)}_
{e_{\mu},\hat{a}}
(v_{1}+\delta ,-\frac{1}{rw})\phi_{\nu}(-v_{2})
\stackrel{\sim}{\varphi}^{(j)}
_{e_{\nu},\hat{a}}(v_{2}+\delta ,-\frac{1}{rw})
|_{F_{l,k}}\\
& &\ \ \ \ =
\sum_{\mu\nu}\phi_{\mu}(-v_{1})\phi_{\nu}(-v_{2})
\stackrel{\sim}{\varphi}^{(i)}_{e_{\mu},\hat{a}+e_{\nu}}
(v_{1}+\delta ,-\frac{1}{rw})
\stackrel{\sim}{\varphi}^{(j)}_{e_{\nu},\hat{a}}
(v_{2}+\delta ,-\frac{1}{rw})|_{F_{l,k}}\\
& &\ \ \ \ =
\sum_{\mu\nu}\sum_{\mu '\nu '}\phi_{\nu '}(v_{2})
\phi_{\mu '}(v_{1})
W'(\hat{\pi}|v_{1}-v_{2},-\frac{1}{rw})
^{\mu\nu}_{\mu '\nu '}
\stackrel{\sim}{\varphi}^{(i)}_{e_{\mu},\hat{a}+e_{\nu}}
(v_{1}+\delta ,-\frac{1}{rw})
\stackrel{\sim}{\varphi}^{(j)}_{e_{\nu},\hat{a}}
(v_{2}+\delta ,-\frac{1}{rw})|_{F_{l,k}}\\
& & \ \ \ = \sum_{i'j'}
\sum_{\mu '\nu '}\phi_{\nu '}(v_{2})
\phi_{\mu '}(v_{1})
\stackrel{\sim}{\varphi}^{i'}_{e_{\mu '},\hat{a}}
(v_{1}+\delta ,-\frac{1}{rw})
\stackrel{\sim}{\varphi}^{j'}_{e_{\nu},\hat{a}+e_{\mu}}
(v_{2}+\delta ,-\frac{1}{rw})
R^{ij}_{i'j'}(v_{1}-v_{2},-\frac{1}{rw})|_{F_{l,k}}
\end{eqnarray*}
\noindent Namely ,we have  the bosonization for vertex operator of 
Zn symmetric Belavin model
\begin{eqnarray}
\Phi^{(i)}(v_{1})\Phi^{(j)}(v_{2})|_{L_{i}}
=\sum_{i'j'}\Phi^{(j')}(v_{2})\Phi^{(i')}(v_{1})
R^{ij}_{i'j'}(v_{1}-v_{2},-\frac{1}{rw})|_{L_{i}}
\end{eqnarray}
\noindent We also can give the bosonization for the Hamiltonian of CTM 
$D^{(i)}$ in Fock space
\begin{eqnarray}
& &D=\sum_{m=1}^{\infty}\sum_{j=1}^{n-1}\frac{[rm]_{x}}{[(r-1)m]_{x}}
\Omega^{j}_{-m}S^{j}_{m}+\frac{1}{2}\sum_{j=1}^{n-1}
P_{\omega_{j}}P_{\alpha_{j}}\\
& & \Omega^{j}_{-m}=\sum^{j}_{k=1}x^{(2k-j-1)m}\beta^{k}_{-m}\ \ ,\ \ 
S^{j}_{m}=x^{-jm}(\beta^{j}_{m}-\beta^{j+1}_{m})\nonumber\\ 
& &D|_{L_{i}}=D^{(i)}|_{L_{i}}\nonumber\\
& &x^{nD}\Phi^{(j)}(v)x^{-nD}=\Phi^{(j)}(v+nw)
\end{eqnarray}
We can also define the dual vertex operators $\Phi^{*}_{\mu}(v)$ through 
the skew-symmetric fusion of n-1s $\Phi(v)$ $^{[10]}$
\begin{eqnarray*}
& &\Phi^{*}_{j}(v)=\sum_{\mu}\overline{\phi}^{*(n-1)}_{\mu}(v-\frac{n}{2})
A^{-1}_{\mu}\varphi^{j}_{e_{\mu},\hat{a}-e_{\mu}}(v,-\frac{1}{rw})\\
& &A_{\mu}=(-1)^{n-1}\frac{x}{(x^{2};x^{2r})(x^{2r-2};x^{2r})}
\prod-{k=1}^{n}\left[ 1+\hat{\pi}_{k\mu}\right]\\
& &\overline{\phi}^{*(n-1)}_{\mu}(v)=\oint\prod^{n-1}_{j=\mu}
\frac{d(x^{2v_{j}})}{2i\pi x^{2v_{j}}}
\eta_{n-1}(v)\xi_{n-1}(v_{n-1})...\xi_{\mu}(v_{\mu})\prod^{n}_{j=\mu+1}
f(v_{j-1}-v_{j},\hat{\pi}_{\mu j})\\
& &{\rm where\ \ we\ \ set\ \ }v_{n}=v \ \ {\rm and} \ \  
|xx^{2v_{j-1}}|<|x^{2v_{j}}|<|x^{-1}x^{2v_{j-1}}|\ \ \ (j=\mu,....,n-1)
\end{eqnarray*}
\noindent Following Eq(c.20) in the Ref.[10] and Eq.(10) ,we have following  
invertibility
\begin{eqnarray}
& &\Phi^{(i)}(v)\Phi^{*}_{j}(v)|_{L_{i}}=c^{-1}_{n}\delta^{i}_{j}
\times id|_{L_{i}}\\
& &c_{n}=x^{\frac{r-1}{r}\frac{n(n-1)}{2n}}\frac{g_{n-1}(x^{n})}
{(x^{2};x^{2r})(x^{2r};x^{2r})^{2n-3}}\nonumber
\end{eqnarray}
Thus the correlation function for Zn symmetric Belavin model can be
described by the following trace functions
\begin{eqnarray}
F^{(i)}(v_{1},....,v_{N})_{i_{1},...,i_{N}}
=\frac{tr_{L_{i}}(x^{nD}\Phi^{(i_{1})}(v_{1}).....\Phi^{(i_{N})}(v_{N})
\Phi^{*}_{i_{N}}(v_{N})\Phi^{*}_{i_{N-1}}(v_{N-1}).....
\Phi^{*}_{i_{1}}(v_{1})}
{tr_{L_{i}}(x^{nD})}
\end{eqnarray}
\noindent Using the cyclic properties of a matrix trace and the relations 
Eq.(34) and Eq.(36), it is easy to derive the difference equations 
which the correlation functions should be satisfied.The difference equations 
for eight-vertex model (n=2) was given by Jimbo et al $^{[20]}$ from the ``
physical picture", and the difference equation for Zn symmetric Belavin model 
(n is generic integer,$2\leq n$) was given by Quano$^{[19]}$ also through 
the ``phyiscal picture".The correlation functions for Zn symmetric 
Belavin model in Eq.(38) can be expressed explicitly in terms of integrals
after carrying out the trace functions of Eq.(38). Therefore ,the trace 
functions of Eq.(38) is the solution to difference equations with elliptic 
R-matrix as its construction coefficient .To directly solve the general 
difference equations of elliptic type is still an open problem$^{[24,27]}$. 
Fortunately, the bosonization method give a system way to solve these 
difference equations.

In the simple case for N=1 ,the trace functions will give the chacter of 
Z-grade spaces $L_{i}$
\begin{eqnarray}
& &tr_{L_{i}}(x^{nD})=\frac{\sum_{\{m_{j}\}\in Z}
x^{\frac{1}{2}\sum_{k=1}^{n-1}
<\omega_{k},\sqrt{\frac{r}{r-1}}l-\sqrt{\frac{r-1}{r}}
(\overline{\Lambda}_{i}+\sum_{j=1}^{n-1}m_{j}\alpha_{j})>
<\alpha_{k},\sqrt{\frac{r}{r-1}}l-\sqrt{\frac{r-1}{r}}
(\overline{\Lambda}_{i}+\sum_{j=1}^{n-1}m_{j}\alpha_{j})>}}
{(x^{2n};x^{2n})^{n-1}}\nonumber\\
\end{eqnarray}

\noindent {\bf Remark:} Actually, the above chacter of space 
$L_{i}$ are the same as that of 
level one integrable represenation of q-deformed affine algebra 
($U_{q}(\hat{sl(n)})$, of couse, it is also equal to that of level one 
representation of affine algebra ($A^{(1)}_{n-1}$)$^{[18]}$.
Thus the space $L_{i}$ 
would be some level one representation of some elliptic deformation 
of affine algebra,
which are not known but many phenomena suggest that it would be existed.
We expect to find this elliptic deformation of affine algebra which would 
play a role of  the symmtric algebra of the elliptic vertex model.

For generic N ,one will encounter the following trace functions
\begin{eqnarray}                        
tr_{L_{i}}(x^{nD}
e^{\sum_{m=1}^{\infty}\sum_{j=1}^{n-1}A^{j}_{m}\beta^{j}_{-m}}
e^{\sum_{m=1}^{\infty}\sum_{j=1}^{n-1}B^{j}_{m}\beta^{j}_{m}}
f^{P_{\gamma}})
\end{eqnarray}
Since the operator $e^{Q_{\alpha}}$ would shift a Fock sector to 
another different sector unless $Q_{\alpha}=0$ , the term of 
$e^{Q_{\alpha}}$ in general no-zero  
trace functions  should be equal to 1. We can caculate the contributions 
in the trace for tensor components (i) oscillators modes and (ii) 
the zero mode separately .The trace over oscillator part can be carry out 
by using the Clavelli-Shapiro technique$^{[25]}$. More explicitly, let us 
introduce other oscillators $\overline{\beta}^{j}_{m}\ \ (j=1,...,n-1)$ 
which commutate with the old ones $\beta^{j}_{m}$. Define the following
operators acting in the tensor product of Fock space of $\beta^{j}_{m}$ 
and that of $\overline{\beta}^{j}_{m}$
\begin{eqnarray*}
& &b^{j}_{m}=\frac{\beta^{j}_{m}\otimes 1}{1-x^{2nm}}+1\otimes 
\overline{\beta}^{j}_{-m}\ \ ,\ \ m>0\\
& &b^{j}_{m}=\beta^{j}_{m}\otimes 1+
\frac{1\otimes\overline{\beta}^{j}_{-m}}{x^{2nm}-1}
\ \ ,\ \ m<0
\end{eqnarray*}
\noindent Now the trace  of some bosonic operator $O(\beta^{j}_{m})$ 
can be expressed in terms of the vacuum expectation value 
$<0|O(b^{j}_{m})|0>$ .Namely,
\begin{eqnarray}
tr(x^{nD}O(\beta^{j}_{m}))=\frac{<0|O(b^{j}_{m})|0>}{(x^{2n};x^{2n})}
\end{eqnarray}
\noindent We denote $<0|O(b^{j}_{m})|0>$ by $<< O(\beta^{j}_{m})>>$ (
we choose the same symbol as that of the Lukaynov's in the Ref.[8]). Due  
to the Wick theorem, the expectation value of a product of exponential 
operators is factorized into the two point functions
\begin{eqnarray}
& &<<\eta_{1}(v_{1})\eta_{1}(v_{2})>>=<<\eta_{n-1}(v_{1})\eta_{n-1}(v_{2})>>
C^{2}_{1}G_{1}(v_{2}-v_{1})\\
& &<<\eta_{1}(v_{1})\eta_{n-1}(v_{2})>>=C^{2}_{1}G_{n-1}(v_{2}-v_{1})\\
& &<<\eta_{1}(v_{1})\xi_{1}(v_{2})>>=C_{1}C_{2} S(v_{2}-v_{1})\\
& &<<\xi_{j}(v_{1})\xi_{j+1}( v_{2})>>=C^{2}_{2} S(v_{2}-v_{1})\\
& &<<\xi_{j}(v_{1})\xi_{j}(v_{2})>>=C^{2}_{2}T(v_{2}-v_{1})\\
& &<<\eta_{n-1}(v_{1})\xi_{n-1}(v_{2})>>=C_{1}C_{2}S(v_{2}-v_{1})\\
& &\{z\}'=(z;x^{2r},x^{2n},x^{2n},x^{2n})\ \ \ 
\rho (\eta_{j})=C_{1}=\frac{\{x^{2+2n}\}'\{x^{2r+4n-2}\}'}
{\{x^{2r+2n}\}'\{x^{4n}\}'}\nonumber\\
& &\rho (\xi_{j})=C_{2}=(x^{2n};x^{2n})\frac{\{x^{2+2n}\}}
{\{x^{2r-2+2n}\}}\nonumber\\
& &G_{1}(v)=\frac{\{x^{2+2v}\}'\{x^{2r+2n-2+2v}\}'\{x^{2+2n-2v}\}'
\{x^{2r+4n-2-2v}\}'}
{\{x^{2r+2v}\}'\{x^{2n+2v}\}'\{x^{2r+2n-2v}\}'\{x^{4n-2v}\}'}\nonumber\\
& &G_{n-1}(v)=\frac{\{x^{n+2v}\}'\{x^{2r+n+2v}\}'\{x^{3n-2v}\}'
\{x^{2r+3n-2v}\}'}
{\{x^{2r+n-2+2v}\}'\{x^{2+n+2v}\}'\{x^{2r+3n-2-2v}\}'\{x^{3n+2-2v}\}'}
\nonumber\\
& &S(v)=\frac{\{x^{2r-1+2v}\}\{x^{2r-1+2n-2v}\}}
{\{x^{1+2v}\}\{x^{1+2n-2v}\}}\nonumber\\
& &T(v)=(x^{2v};x^{2n})(x^{2n-2v};x^{2n})
\frac{\{x^{2+2v}\}\{x^{2+2n-2v}\}}
{\{x^{2r-2+2v}\}\{x^{2r-2+2n-2v}\}}\nonumber
\end{eqnarray}
\noindent For all other combinations of $\eta_{1}(v)\ \ ,\eta_{n-1}(v)\ \ ,
\xi_{j}(v)$, we have $<<XY>>=\rho (X) \rho (Y)$ and $\rho (\eta_{j})=C_{1}$ 
$\rho (\xi_{j})=C_{2}$.
\section{Discussions}
In this paper we construct the bosonic realization of vertex operators 
for Zn symmetric Belavin model in the Fock spaces $\oplus_{i=1}^{n-1}L_{i}$. 
In fact,they are the intertwiner operators among spaces $L_{i}$ ,and  
$L_{i}$ would be the level one representaion space of some 
deformed affine algebra 
(should not be q-deformed affine algebra).Morever,we guess that 
this deformed algebra would be some elliptic deformed affine algebra: 
for the case of eight-vertex model (n=2) it  would be the elliptic 
algebra $A_{q,p}(\hat{sl_{2}})$ $^{[23]}$; for the case of $2<n$,
it would be some 
elliptic algebra which is generalization of $A_{q,p}(\hat{sl_{2}})$ for   
high rank n. We will present the results of this kind algebra in the further 
paper.

We only consider the type I vertex operators$^{[3]}$. We can further 
construct the bosonization for type II vertex operators of Zn symmetric 
Belavin model .

\end{document}